\documentstyle[epsfig,mncite]{mn}
\begin{document}
\newcommand\msun   {M$_{\odot}$}
\LARGE
\normalsize
\title{The mass of the neutron star in the low--mass X--ray binary
2A~1822--371 $^*$}

\author[P.G. Jonker et al.]
{P.G. Jonker$^{1}$\thanks{email : peterj@ast.cam.ac.uk \newline $^*$Based on
observations made with ESO Telescopes at the Paranal Observatories
under programme ID 267.D--5683},
M. van der Klis$^2$,
P.J. Groot$^{3,4}$\\
$^1$Institute of Astronomy, Madingley Road, CB3 0HA, Cambridge\\
$^2$Astronomical Institute ``Anton Pannekoek'',
University of Amsterdam, Kruislaan 403, 1098 SJ Amsterdam\\
$^3$Harvard--Smithsonian Center for Astrophysics, 60 Garden
Street, Cambridge, MA 02138, USA\\
$^4$Department of Astrophysics, University of Nijmegen, P.O.Box 9010,
Nijmegen, The Netherlands\\ } \maketitle

\begin{abstract}
\noindent
Using phase resolved spectroscopic observations obtained with the
Ultraviolet and Visual Echelle Spectrograph on ESO's Kueyen Very Large
Telescope supplemented by spectroscopic observations obtained with the
Boller and Chivens Spectrograph on the Walter Baade Magellan
telescope, we found sinusoidal radial--velocity variations with a
semi--amplitude 327$\pm$17 km/s. From previous observations and from
the fact that the epoch of minimum velocity arrived early with respect
to the epoch calculated from pulse timing we know that the companion
star is suffering from irradiation. Since we most likely observed
primarily the side of the companion star facing the observer at phase
$\sim$0.75 the velocity quoted above is not the true radial velocity
semi--amplitude of the companion star. Assuming a uniform contribution
to the line profile from this hemisphere yields a radial velocity
semi--amplitude of 280$\pm$26 km s$^{-1}$ for a systemic velocity of
54$\pm$24 km s$^{-1}$; if the contribution is instead weighted
somewhat more towards the side of the companion facing the X--ray
source then the true semi--amplitude is larger than this value.
Together with the well constrained inclination (81$^\circ<i<84^\circ$)
and the mass--function determined from pulse--timing analysis
(2.03$\pm$0.03$\times10^{-2}$\,M$_\odot$), we derive a lower limit to
the mass of the neutron star and to that of the companion star of
0.97$\pm$0.24\,M$_\odot$ and 0.33$\pm$0.05\,M$_\odot$, respectively
(1\,$\sigma$; including uncertainties in the inclination). We briefly
discuss other aspects of the spectrum and the implications of our
findings.

\end{abstract}

\begin{keywords} stars: individual (2A~1822--371) --- stars: neutron
--- X-rays: stars --- techniques: radial velocities
\end{keywords}

\section{Introduction}
\label{intro}

Low--mass X--ray binaries (LMXBs) are generally old ($>10^8$\,yr)
binary systems in which a low--mass star ($\la1\,M_{\sun}$) transfers
matter to a neutron star or a black hole. The neutron star LMXBs are
thought to be progenitors of millisecond radio pulsars. Due to
accretion of matter and decay of the magnetic field during the
LMXB--phase the neutron star spins--up to millisecond periods
(\pcite{radsri1982}; \pcite{1982Natur.300..728A}; see
\pcite{bhatta1995} for a review). To date, only eight out of several
hundred LMXBs, among which three with millisecond periods, are known
to show pulsations. Measuring Doppler delays of the pulse arrival
times allowed for an accurate measurement of both the orbital period
and the size of the orbit of the neutron star in six cases (Her~X--1,
which is in fact an intermediate--mass X--ray binary,
\pcite{1972ApJ...174L.143T}; GRO~J1744--28,
\pcite{1996Natur.381..291F}; SAX~J1808.4--3658,
\pcite{1998Natur.394..346C}; 2A~1822--371,
\pcite{2001ApJ...553L..43J}; XTE~J1751--305,
\pcite{2002IAUC.7867....1M}; XTE~J0929--314,
\pcite{2002IAUC.7900....2G}). The orbit of the companion star can be
determined by observing periodic shifts in the central wavelengths of
(absorption) lines in its stellar spectrum. If the orbital inclination
is known, one can solve for the masses of the stellar components.

The initial mass of newly formed neutron stars is highly uncertain but
theoretical calculations show that they most likely fall in a small
range around 1.32\,$M_{\odot}$ (\pcite{1996ApJ...457..834T};
\pcite{2001ApJ...554..548F} derived that 81--96 percent of the initial
neutron star masses fall in the range of 1.2--1.6\,$M_{\odot}$).
Theories on the equation of state (EoS) of neutron--star matter at
supranuclear density provide a firm upper limit on the neutron star
mass which is different for each EoS. Therefore, measuring a high mass
for even one neutron star would imply the firm rejection of many
proposed EoS (see discussion by van Paradijs \& McClintock 1995). The
measured masses of radio pulsars, including millisecond radio pulsars,
are consistent with 1.4$M_\odot$ (Thorsett \& Chakrabarty 1999).  Some
theories explaining the presence of kHz quasi--periodic oscillations
in accreting non--pulsating LMXBs (see van der Klis 2000 for a review)
suggest a neutron star mass of $\sim 2 M_\odot$.

The lightcurve of 2A~1822--371 shows clear signs of orbital modulation
in both the X--ray and optical bands (e.g. partial eclipses and a
sinusoidal modulation in optical and X--rays;
\pcite{1981ApJ...247..994W}; \pcite{1979IAUC.3406....1S};
\pcite{1980ApJ...242L.109M}) with an orbital period of 5.57 hours. The
implied inclination of 2A~1822--371 is $i=82^\circ-87^\circ$
(\pcite{1989MNRAS.239..715H}). \scite{2001MNRAS.320..249H} used {\it
  ASCA} and {\it RXTE} data to model the lightcurve; they found
$81^\circ<i<84^\circ$. Recently, 0.59~s pulsations were discovered
from this system and from pulse arrival time delay measurements it was
found that the neutron star in 2A~1822--371 has an {\sl a}sin$i$ of
1.006~lightseconds (\pcite{2001ApJ...553L..43J}). From eclipse and
pulse timing an accurate ephemeris is known (\pcite{parmaretal2001};
\pcite{2001ApJ...553L..43J}). Clearly, spectroscopic measurements of
the radial velocity curve of the companion are indicated at this stage
in order to check on the mass of the neutron star.

Cowley, Crampton, \& Hutchings (1982) detected weak absorption lines
in the spectrum of 2A~1822--371 (H$\delta$, H$\gamma$, and some HeI
absorption lines) but due to the lack of signal--to--noise a radial
velocity could not be measured.  Later, Harlaftis, Charles, \& Horne
(1997) measured the radial velocity of the companion in 4U~1822--37
using the He~I absorption line at 5875.966\AA. The presence of an
extra absorption component in the same part of the spectrum and the
relatively low resolution ($\sim$75 km s$^{-1}$ for H$\alpha$) which
kept them from resolving the two components, yielded a lower limit on
the radial velocity of the companion star and mass of the neutron star
of 225 km s$^{-1}$ and $0.6^{+1.0}_{-0.3}$M$_\odot$, respectively.

In this Letter we report on the spectroscopic observations of
2A~1822--371 obtained with the Ultraviolet and Visible Echelle
Spectrograph on ESO's 8.2-m Kueyen Very Large Telescope and on
spectroscopic observations obtained with the Boller and Chivens
Spectrograph on the 6.5-m Walter Baade Magellan Telescope.

\section{Observations and analysis}
\label{analysis}
2A~1822--371 was observed with the Ultraviolet and Visible Echelle
Spectrograph (UVES) mounted on the Kueyen Very Large Telescope (VLT)
from UT 00:51--07:10 July 20, 2001 (HJD 2452110.546--2452110.808). Due
to a technical problem with the telescope no observations were
obtained from UT 05:27--06:20; this led to a gap in the coverage of
the binary orbit from phase 0.38 to 0.50. The integration time was
900~s. UVES was operated in its 390$+$564 standard mode with a slit
width of 1 arcsecond. This yielded spectra with a resolution of
$\sim$0.1\AA\,per pixel. The night was photometric with a seeing less
than 1 arcsecond. Thorium--Argon lamp spectra were obtained after each
observing block of one hour.

Spectroscopic observations with the Walter Baade Magellan Telescope
were obtained on June 30 (UT 01:48--09:28; HJD
2452090.584--2452090.896), July 1 (UT 03:17--09:39; HJD
2452091.646--2452091.904), and July 2 (UT 03:22--09:25; HJD
2452092.649--2452092.894) 2001 using the Boller and Chivens
Spectrograph. The slit--length and width were 72 and 0.75 arcseconds,
respectively, providing a resolution of 0.9\AA\, per pixel. The
integration time was 600~s. Helium--Argon lamp spectra were
obtained approximately every hour. The spectra were bias--subtracted,
flatfielded, extracted using the {\sc echelle} (VLT) and {\sc specred}
(Magellan) reduction packages and fitted with routines in the {\sc
stsdas} package in {\sc IRAF}\footnote{{\sc IRAF} is distributed by the
National Optical Astronomy Observatories}.

UVES spectra were extracted from 3700--4500\AA, 4640--5580\AA, and
5670--6660\AA. The rms scatter of the wavelength calibration obtained
by applying a fourth order polynomial fit to the wavelengths of more
than 50 lamp lines was $\sim0.003$\AA. The source spectra were
rebinned to a resolution of $\sim$0.3\AA. The normalisation of the
spectra occured in three stages. First, for each order, the average
blaze profile was determined by fitting a cubic spline to the average
spectrum of all observations combined. The regions where lines were
known to be present from previous spectroscopic observations of the
source (see Section~\ref{intro}) were masked during the fit. Second,
each order, for each spectrum, was divided by this fit. Finally, all
orders were combined and scaled to a mean level of 1.

The Magellan spectra cover 5310\AA--6300\AA.  The wavelength
calibration was obtained from a second order polynomial fit to 12 lamp
lines (rms of $\sim$0.05\AA). A second order cubic spline fit to the
continuum was used to take out the instrumental profile and normalise
the spectrum.

Using the ephemeris provided by the pulse timing
(\pcite{2001ApJ...553L..43J}) we determined the phase at the centre of
each spectrum. The Magellan observations falling in the same phase bin
(bin--width 0.03) were averaged to increase the signal--to--noise. The
error in the phase is dominated by the integration time used to obtain
the spectrum and/or the spread in phase when averaging the Magellan
data, and not by the propagation of the ephemeris to the date of the
observation. Phase zero is defined as superior conjunction of the
neutron star (X--ray eclipse).

\section{Results}
\label{results}
\begin{figure*}
\leavevmode{\psfig{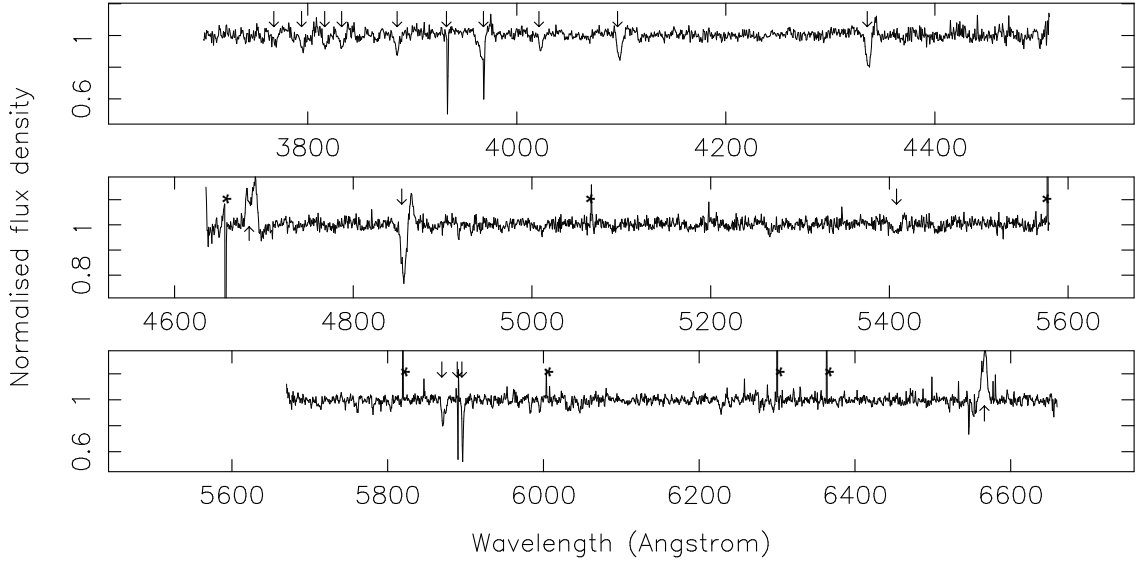}}\caption{Total
spectrum at phase 0.67$\pm$0.02 rebinned to a wavelength resolution of
0.6\AA\,. The H$\iota$, H$\theta$, H$\eta$, H$\zeta$, H$\epsilon$,
H$\delta$, H$\gamma$, H$\beta$ hydrogen Balmer lines, the
double--peaked He~II and H$\alpha$ emission lines, and the He~I
absorption lines at 4026.357\AA\, and 5875.966\AA\, are clearly
visible. The most prominent lines have been indicated with an arrow,
whereas noise spikes due to residual cosmics or CCD imperfections have
been indicated with a star.}
\label{spectra}
\end{figure*}

In Figure~\ref{spectra} we show as an example the full UVES spectrum
at orbital phase 0.67$\pm$0.02 rebinned to a wavelength resolution of
0.6\AA. Since the spectra are varying strongly as a function of
orbital phase, we were not able to cross--correlate the spectra in
order to determine the radial velocity. Instead, we fitted Gaussians
to the line profiles and compared the central wavelength with the rest
wavelength of the line. Fitting was done using a downhill simplex
method ({\sc amoeba}; \pcite{1992nrfa.book.....P}) and errors on the
parameters were determined using an independent Monte Carlo resampling
technique.

\begin{table*}
\caption{Overview of the detected lines. In the last column the
range in phase over which the line was predominantly detected is given.}
\label{lines}
\begin{center}

\begin{tabular}{lccc}
\hline

Element & Rest wavelength (\AA) & Absorption (A)/emission (E) &
Present at phases\\
\hline
\hline
H$\iota$ & 3770.630 & A & 0.55-1\\
H$\theta$ & 3797.898 & A & 0.55-1\\
He~I    & 3809.10  & E & 0.2-0.35, 0.65-1\\
H$\eta$ & 3835.384 & A & 0.55-1\\
He~I    & 3888.60  & E & 0.9-1, 0-0.1\\
H$\zeta$ & 3889.049 & A $+$ E (sometimes double) & 0.55-1\\
Ca~II   & 3933.663 & A interstellar & 0-1\\
Ca~II   & 3968.468 & A interstellar & 0-1\\
H$\epsilon$ & 3970.072 & A $+$ E (sometimes double) & 0.5-1\\
He~I    & 4026.357 & A & 0.3-1\\
H$\delta$ & 4101.734 & A $+$ E (sometimes double) & 0.55-1\\
H$\gamma$& 4340.464 & A $+$ E (sometimes double) & 0.55-1\\
He~II   & 4685.71  & E double & 0-1\\
H$\beta$& 4861.325 & A $+$ E (sometimes double) & 0.5-1\\
He~II   & 5411.53  & E & 0.95-1, 0-0.1\\
He~I    & 5875.966 & A & 0.5-1 \\
Na~D    & 5889.951 & A interstellar & 0-1\\
Na~D    & 5895.924 & A interstellar & 0-1\\
H$\alpha$& 6562.80  & E double & 0-1\\

\end{tabular}
\end{center}

\end{table*}

\begin{figure*}
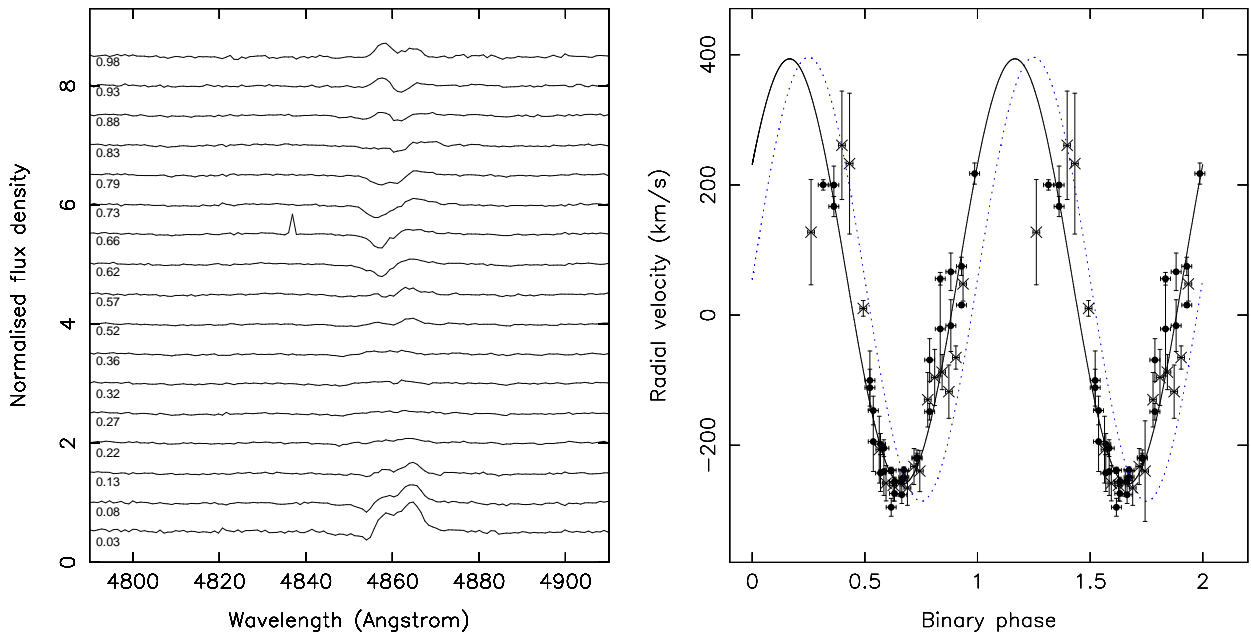

\leavevmode{\psfig{file=4861.ps,width=8cm}
\quad
\psfig{file=radial.ps,width=8cm}}
\caption{{\it Left:} The changing profile of H$\beta$ as a function of
  binary orbital phase. The phase is given by the number on the left
  in the plot. The spectra have been shifted by 0.5 units in flux for
  clarity. {\it Right:} The radial velocity as a function of binary
  orbital phase determined from the He~I 4026.357 \AA\,and
  5875.966 \AA\,absorption lines. The dots are the measurements using
  the VLT data and the crosses those of the Magellan data. For clarity
  two cycles have been plotted. The best--fit sinusoid as well as the
  fit with the phase of the sinusoid fixed to the value expected from
  the pulse--timing ephemeris are overplotted (solid line and the
  dotted line, respectively).}
\label{4861}
\label{vrad}
\end{figure*}

We detected several lines of the hydrogen Balmer series (among which
the double--peaked H$\alpha$ emission line). Their profiles varied as
a function of orbital phase. To show the variablity of the hydrogen
absorption lines in more detail we plotted a blow--up of the region of
H$\beta$ as a function of phase in Figure~\ref{4861} ({\it left
  panel}). H$\beta$ changed from a double--peaked emission line into a
strong absorption line with a single emission line on the red side,
and back into a strong double--peaked emission line with a weak
absorption to the blue side. This pattern is detected in the two
strongest hydrogen Balmer absorption lines (H$\beta$ and H$\gamma$).
Hints of the presence of the same pattern were found for the weaker
hydrogen Balmer lines below 4000\AA\,as well, but due to the limited
signal--to--noise the presence or absence of such components could not
always be determined. These double--peaked lines are most likely formed
in the optically thin outer parts of the accretion disk.

Several He~I lines were detected (e.g. the He~I 5875.966\AA\,
absorption line also reported by \pcite{1997MNRAS.285..673H}, the He~I
4026.357\AA\, absorption line and the He~I 3809.10\AA\, and
3888.60\AA\, emission lines). The He~I lines did not show evidence for
changes in line--profile such as those found for the hydrogen Balmer
lines, although since the lines are often weak we cannot always
exclude that the lines are composites of lines originating both in the
accretion disk and in the secondary star. The Bowen blend, detected in
2A~1822--371 by \scite{1982ApJ...262..253M} was not clearly detected
because it fell close to the edge of the CCD. Double--peaked He~II
emission lines with rest wavelength of 4685.71\AA\,and 5411.53\AA\,
were also detected.  Their strength and profile changed strongly as a
function of orbital phase, similar to the double--peaked H$\beta$ and
H$\gamma$ lines. A list of lines which were clearly detected is given
in Table~\ref{lines}.

For our radial velocity study we only used the He~I absorption lines
at 4026.357\AA\, and 5875.966\AA\,(see Figure~\ref{vrad}; {\it right
  panel}; dots are measurements using VLT data and crosses are
measurements using Magellan data. Note that the two sets of
measurements are completely consistent without any evidence for a
systematic offset.) We excluded the hydrogen Balmer lines since we
could not always unambiguously separate the emission component(s) from
the absorption component. When the hydrogen absorption and emission
components could unambiguously be separated the velocities determined
from the hydrogen absorption lines were consistent with those of the
He~I absorption lines. The He~I lines were always detected at phases
0.3 to 0.95. We fitted a sinusoid to the measured radial--velocity
curve using the orthogonal distance regression (ODR) technique to
minimize the $\chi^2$ (the reported errors indicate the 1$\sigma$
uncertainties throughout this paper).

Fixing the orbital period and phase to the values predicted from pulse
timing did not result in a good fit ($\chi^2=436$ for 45 degrees of
freedom). Inspection of the fit reveals that the phase of minimal
velocity occurs earlier than expected on the basis of the orbital
ephemeris (see Figure~\ref{vrad}; {\it right panel} dotted line).
Such effects have been noticed before in radial velocity curves of
several Cataclysmic Variables (CVs; e.g.  \pcite{1992MNRAS.257..476D})
and are thought to be caused by asymmetric heating of the companion
star; i.e.  the centre of light is offset from the centre of mass of
the secondary, and not on the binary axis. Since one side of the
companion star contributes disproportionally, the rotational
broadening of the absorption profile is not uniform; this leads to a
systematic velocity shift (\pcite{1988ApJ...324..411W}).  We will
come back to this in the Discussion. This heating effect can be
accommodated by allowing for a phase shift in the sinusoid or by
introducing a small apparent eccentricity
(\pcite{1992MNRAS.257..476D}).

Leaving the phase of the sinusoid as a free parameter improved the fit
dramatically ($\chi^2=90$ for 44 degrees of freedom; see
Figure~\ref{vrad}, {\it right panel}, solid line; the phase shifts
from 0.0 to --0.08$\pm$0.01). This yields a systemic velocity,
v$_{sys}$, of 67$\pm$15 km s$^{-1}$ and a radial velocity
semi--amplitude, K, of 327$\pm$17 km s$^{-1}$. We note that the
v$_{sys}$ and K are correlated in the fit since the absorption line
measurements are predominantly located at phases 0.5 to 1. After
\scite{1992MNRAS.257..476D} we tried to improve the fit by including
the second harmonic of the sinusoid to accommodate a small, apparent,
eccentricity. The fit with two sinusoids was not significantly better
(cf. $\chi^2=82$ for 42 degrees of freedom for a two--sinusoids fit
vs. $\chi^2=90$ for 44 degrees of freedom for a single sinusoid fit).

\section{Discussion}

We determined the radial velocity of the companion star of
2A~1822--371 using VLT and Magellan spectroscopic observations.
Fitting a sinusoid to radial velocity measurements using two He~I
stellar absorption lines a semi--amplitude of 327$\pm$17 km s$^{-1}$
was found, with a best fit systemic velocity of 67$\pm$15 km s$^{-1}$.
We could combine this with the velocity amplitude of the neutron star
(derived from pulse--timing analysis; \pcite{2001ApJ...553L..43J}) to
derive a neutron and companion star mass. However, we found that the
epoch of minimum velocity occurs early with respect to the epoch
calculated from the ephemeris derived from pulse timing. Together with
the fact that the absorption lines were predominantly detected around
phase 0.75 this is strong evidence that the companion star suffers
from asymmetric heating (see also \pcite{1992MNRAS.257..476D}; that
2A~1822--371 suffers from X--ray heating was also shown by
\pcite{1982ApJ...255..596C} and \pcite{1997MNRAS.285..673H}).
Therefore, it is likely that the centre--of--light of the companion
star does not coincide with its centre of mass, nor with the binary
axis.

The effect this has on the determined radial velocity amplitude
depends on what part of the companion star is observed and whether as
assumed here and below the companion star is in co--rotation. In
several CVs absorption lines from the non--irradiated side of the
companion star, the side pointing towards the observer at phase 0 (for
an inclination, $i$, of 90$^\circ$) are observed
(\pcite{1988ApJ...324..411W}). In that case the measured radial
velocity amplitude is an upper limit to the true amplitude. If we were
to observe the irradiated side in such a case the determined radial
velocity amplitude would be a lower limit. Furthermore, the minimum
and maximum of the radial velocity curve would both move closer to
phase 0.5 whereas the velocities at phase 0 and 0.5 should be the same
(and they should be zero).  However, from Figure~\ref{vrad} ({\it
  right panel}) it is clear that in case of 2A~1822--371 these
velocities are unequal and thus this model is not applicable here.

The distribution of points in Figure~\ref{vrad} ({\it right panel})
indicates that the leading side of the companion star was observed
predominantly, i.e., the side pointing towards the observer at phase
0.75 ($i=90^\circ$). If we assume a uniform contribution from this
leading hemisphere to the line profiles, then, due to the rotation of
the companion star, the velocities at phase 0 and 0.5 will be red and
blue shifted, respectively, but the magnitude of the shift will be the
same. The velocity measured at phase 0.25 and 0.75 with respect to the
average of the velocities at phase 0 and 0.5 (which is the systemic
velocity) represents the true radial velocity amplitude in such a
model.  Since only part of the sinusoid is observed in case of
2A~1822--371 the radial velocity semi--amplitude is calculated from
the difference between the average of the velocities at phases 0 and
0.5, and the velocity at 0.75 only.  Using this model we derive a
systemic velocity of 54$\pm$24 km s$^{-1}$ and a radial velocity
semi--amplitude of 280$\pm$26 km s$^{-1}$. Combining this with the
velocity amplitude of the neutron star a neutron star mass of
0.97$\pm$0.24\,M$_\odot$\,is found (1\,$\sigma$; the uncertainty in
the system inclination, $i=82.5^\circ\pm1.5^\circ$, was included; for
the determination of the inclination see references given in the
Introduction). The companion star mass is 0.33$\pm$0.05\,M$_\odot$ for
$i=82.5^\circ\pm1.5^\circ$. 

If instead the orientation of the hemisphere we observe is, as the
data suggest, somewhat more towards the side facing the X--ray source,
then these quoted values of K and the masses are likely to be
underestimates. Clearly, phase--resolved observations detecting lines
from the cold hemisphere of the companion star are needed in order to
investigate the systematic effects further. We note however that
radial velocity measurements using lines from the non--irradiated side
only would lead to an upper limit on the mass of the neutron star.
\scite{1999ApJ...512..288T} determined that the masses of neutron
stars in radio pulsar binaries are consistent with a Gaussian
distribution around M$=1.35\pm0.04$\,M$_\odot$. The lower limit to the
mass of the neutron star in 2A~1822--371 is consistent with this
within 2 $\sigma$.

\scite{1988MNRAS.233..285N} observed several He~I absorption lines
among which the two we observed in 2A~1822--371 in the spectrum of the
source AC~211 in M~15 (the source is thought to be in the class of
Accretion Disk Corona, ADC, sources of which 2A~1822--371 is the prototype;
\pcite{1981ApJ...247..994W}). They argue that the He~I absorption in
AC~211 is due to disk edge material, which when illuminated by the hot
inner part of the accretion disk gives rise to the He~I absorption
lines.  Unexplained in such a scenario is the fact that even at
orbital phases where the inner part of the accretion disk is not
observed through localized disk edge material (e.g. at phases
0.2--0.7) still absorption lines are observed. Furthermore, they argue
that the radial velocity measurements using these lines in AC~211
reflect the orbital motion of the compact object. In case of
2A~1822--371 we know from the pulse timing measurements of the orbit
of the neutron star that this is not the case.

As for several CVs (most notably U~Gem; \pcite{1992MNRAS.257..476D}),
the observed, irradiated, side of the companion star seems not centred
towards the neutron star or white dwarf in the case of CVs. The reason
for this is unclear. It has been speculated that the hot spot plays an
important role in heating the companion star in CVs
(\pcite{1992MNRAS.257..476D}). However, as for CVs
(\pcite{smithrc1995}) this seems unlikely in case of 2A~1822--371
since the centre of the accretion disk and the neutron star in LMXBs
such as 2A~1822--371 should be much brighter than the hot spot. Even
though shielding of the companion star by the accretion disk could be
important the large ADC should circumvent shielding at least
partially. Modelling irradiation--induced motions in the secondary
stars in CVs \scite{1995MNRAS.275...31M} showed that the heated
material is deflected towards the leading hemisphere as a result of
the Coriolis force; this is in accordance with our findings (see
\pcite{smithrc1995} for a review).

Comparing our data with previous spectroscopic observations
(\pcite{1980ApJ...241.1148C}; \pcite{1982ApJ...255..596C};
\pcite{1982MNRAS.200..793M}; \pcite{1997MNRAS.285..673H}) we note that
\scite{1997MNRAS.285..673H} measured the He~I 5875.966\AA\,line to be
present over the whole orbit. Furthermore, a separate absorption
component arose near phase 0.0--0.2. In our higher resolution UVES
data we did not find evidence for this extra component, nor were we
able to measure the He~I line at all phases.
\scite{1982MNRAS.200..793M} already pointed out that the spectra
showed evidence for long term variability. With our new
high--resolution UVES spectra we can settle the issue of the presence
of emission or absorption lines in the blue part of the spectrum (see
Discussion in \pcite{1982MNRAS.200..793M}); we now know that at
certain phases both (He~I) emission and hydrogen Balmer absorption
lines are present. 

As was noted before, for a neutron star mass of $\sim$1.4\,M$_\odot$
the companion star is undermassive, i.e. less massive than a
main--sequence companion star filling its Roche--lobe in a 5.57 hour
orbit (main--sequence mass $\sim$0.62\,M$_\odot$; e.g.
\pcite{1982MNRAS.200..793M}; \pcite{1982ApJ...255..596C};
\pcite{2001ApJ...553L..43J}). Thus the density of the star, like that
of evolved stars, is lower than the density of a main--sequence star.
The companion stars in some CVs were also shown to be undermassive.
It is believed that the companion star in those CVs is out of thermal
equilibrium (see \pcite{1998A&A...339..518B}). However, if the companion star
in 2A~1822--371 is evolved this would mean that it was considerably
more massive than 0.4\,M$_\odot$ at the onset of mass transfer since
such a star would not have evolved in a Hubble time. Alternatively, a
scenario in which the companion star was brought into contact with its
Roche lobe by transfer of angular momentum due to gravitational
radiation can be envisaged. However, such losses are not sufficient to
sustain the mass transfer rate of $\sim2\times10^{-9}$ M$_\odot {\rm
  yr}^{-1}$ which is needed to explain the intrinsic X--ray luminosity
(L$_X\sim2-4\times10^{37}{\rm erg\,s^{-1}}$;
\pcite{2001ApJ...553L..43J}). Since X--ray irradiation of the
companion star was shown to heat the star, irradiation induced
mass--transfer (\pcite{1989ApJ...343..292R}) could in principle also
help to keep mass--transfer going. However, due to the presence and
size of a bulge on the accretion disk edge a large fraction of the
companion star will be shielded and the irradiating flux will be
reduced, making it unlikely that a mass transfer rate of
$\sim2\times10^{-9}$ M$_\odot {\rm yr}^{-1}$ is induced by X--ray
irradiation. In addition to gravitational radiation and X--ray
irradiation angular momentum transferred by magnetic braking
(\pcite{1981A&A...100L...7V}) could explain the mass--loss rate
(although it was found that at present the orbit is expanding;
\pcite{pawhgi1986}). We conclude therefore that the companion star in
2A~1822--371 is either out of thermal equilibrium (possibly due to the
high mass loss rate and effects of X--ray heating), or evolved and has
lost more than $\sim$0.4\,M$_\odot$ (this would make the companion
star $\sim$0.8\,M$_\odot$ at the start of mass transfer; such a star
could just have evolved in a Hubble time), or an interplay between
angular momentum losses caused by magnetic braking and gravitational
radiation supplemented by mass--loss driven by X--ray irradiation is
driving the evolution of this LMXB.

\section*{Acknowledgments} 
\noindent 
PGJ is supported by EC Marie Curie Fellowship HPMF--CT--2001--01308.
MK is supported in part by a Netherlands Organization for Scientific
Research (NWO) grant. PGJ would like to thank Gijs Nelemans for
numerous useful discussions. The authors would like to thank the
referee for his/her comments and suggestions which helped improve the
paper.

\end{document}